\documentclass[manuscript]{aastex}

\newcommand{\be}{\begin{equation}}
\newcommand{\ee}{\end{equation}}
\newcommand{\beq}{\begin{eqnarray}}
\newcommand{\eeq}{\end{eqnarray}}
\newcommand{\unit}[1]{\ensuremath{\, \mathrm{#1}}}

\newcommand{\vect}[1]{{\bf #1}}

\shorttitle{Spectroscopy related to a sunquake}
\shortauthors{Matthews et al.}

\begin{document}


\title{Spectroscopic signatures related to a sunquake}


\author{S. A. Matthews\altaffilmark{1}, L.K. Harra\altaffilmark{1}}
\affil{UCL Mullard Space Science Laboratory, Holmbury St. Mary, Dorking, Surrey RH5 6NT, U.K.}

\author{S. Zharkov\altaffilmark{2}}
\affil{Department of Mathematics and Physics, University of Hull, Hull, U.K.}

\author{L.M. Green\altaffilmark{1}}
\affil{UCL Mullard Space Science Laboratory, Holmbury St. Mary, Dorking, Surrey RH5 6NT, U.K.}
\email{sarah.matthews@ucl.ac.uk}



\begin{abstract}
The presence of flare related acoustic emission (sunquakes) in some flares, and only in specific locations within the flaring environment, represents a severe challenge to our current understanding of flare energy transport processes. In an attempt to contribute to understanding the origins of sunquakes we present a comparison of new spectral observations from Hinode's EUV imaging Spectrometer (EIS) and the Interface Region Imaging Spectrograph (IRIS) of the chromosphere, transition region and corona above a sunquake, and compare them to the spectra observed in a part of the flaring region with no acoustic signature. Evidence for the sunquake is determined using both time-distance and acoustic holography methods, and we find that, unlike many previous sunquake detections, the signal is rather dispersed, but that the time-distance and 6 and 7 mHz sources converge at the same spatial location. We also see some evidence for different evolution at different frequencies, with an earlier peak at 7 mHz than at 6 mHz. Using EIS and IRIS spectroscopic measurements we find that in this location at the time of the 7 mHz peak the spectral emission is significantly more intense, shows larger velocity shifts and substantially broader profiles than in the location with no sunquake, and that there is a good correlation between blue-shifted, hot coronal, hard X-ray (HXR) and red-shifted chromospheric emission, consistent with the idea of a strong downward motion driven by rapid heating by non-thermal electrons and the formation of chromospheric shocks. Exploiting the diagnostic potential of the Mg II triplet lines, we also find evidence for a single, large temperature increase deep in the atmosphere, consistent with this scenario. 
The time of the 6 mHz and time-distance peak signal coincides with a secondary peak in the energy release process, but in this case we find no evidence of HXR emission in the quake location, but very broad spectral lines, strongly shifted to the red, indicating the possible presence of a significant flux of downward propagating Alfv\'en waves. 

\end{abstract}


\keywords{}



\section{Introduction}

In the 'standard' 2D eruptive flare paradigm (for a review see \cite{shibata11}) energy is released via magnetic reconnection in the solar corona where it is converted to heat, mass motions, and the kinetic energy of accelerated particles, and is subsequently transported both out into the heliosphere and down into the deeper solar atmosphere. Observations of the subsequent response of the plasma to this energy release do, in some aspects (e.g the appearance of flare loops), appear consistent with expectations based on this paradigm, but it is clearly unable to account fully for all the observed elements of the flare process. 

One area that represents a significant challenge for the 'standard' model is the observation of localized acoustic disturbances in response to flaring activity. These so-called sunquakes \citep{kz1998} arguably represent one of the most severe constraints for energy transport through the solar atmosphere, despite the relatively small fraction of total flare energy involved ($\leq$ 10$^{-3}$, \citet{DL2005}). As \cite{judge14} point out, if we assume (as the standard model tells us) that the flare energy is released in the corona, then in order to drive an acoustic disturbance in the solar interior, the energy must propagate through nine pressure scale heights. Since acoustic enhancements are also not detected for every flare, and only in certain locations within the flaring active region when they do occur, understanding their origins can also provide new insights into the effects of environmental (e.g. plasma, magnetic field) conditions on flare energy transport. 

The first sunquake was observed by \cite{kz1998}, and most subsequent studies have indicated that the sunquake originates in the impulsive phase of the flare and is generally well-correlated with the locations of enhanced white-light and hard X-ray emission \citep{Zharkova07, DL2005}. This observed correlation has led to suggestions that chromospheric shocks, which arise as the result of the pressure transients driven by the hydrodynamic response of the ambient plasma to the precipitation of energetic particles into the chromosphere; pressure transients that are related to back-warming of the photosphere by enhanced chromospheric radiation, (e.g. \cite{DL2005}); and Lorentz force transients that occur whilst the coronal field is being reconfigured (e.g. \cite{Hudson_etal12}), may be potential drivers of the acoustic emission. However, the launch of SDO, coupled with now finely-honed detection techniques \citep{ZH11}, has led to a series of sunquake observations that show significantly weaker correlations between the location of the acoustic sources and the areas of strongest particle precipitation and enhanced white-light emission, as well as a hitherto undiscovered connection to the ends of the erupting flux rope \citep{pedram12, Zharkov_etal11, Matthews11, Zharkov2013a, Zharkov2013b} suggesting that we are still a long way from understanding the mechanism(s) responsible for producing these important events. Indeed, recent work by \cite{judge14} reports unique and important observations in the infra-red (IR) during a sunquake that constrains the depth of the flare heating to be in the photosphere (100$\pm$100 km), but also demonstrates that the level of non-magnetic and magnetic energy fluxes  consistent with their observations are too low to drive the acoustic transient. In this work we examine the same flare that \citet{judge14} analysed with a focus on complementary spectroscopic observations of the corona, transition region and the chromosphere above the sunquake. We compare spectra recorded by EIS on Hinode \citep{culhane07} and IRIS \citep{depontieu14} in the location above the quake with spectra above a region in the Northern flare ribbon that has no acoustic enhancement in order to look for signatures of downward propagating energy in these regions and any differences that may shed additional light on the sunquake origin.

\section{Observations and Data Reduction}

On 29 March 2014 an X1 flare (SOL2014-03-29T17:48) occurred in active region NOAA 12017. This was an exceptionally well-covered event, observed from space by Hinode; the Solar Dynamics Observatory (SDO); IRIS,  and the Reuven Ramaty High Energy Solar Spectroscopic Imager (RHESSI) \citep{lin02}, as well as from the ground by the FIRS (Facility Infrared Spectrometer) and Imaging BIdimensional Spectrometer (IBIS) \citep{cavallini06} at the Dunn Solar Telescope (DST). The photospheric signatures of this event, and their relation to the sunquake, have been extensively studied by \cite{judge14}, and the work we present here is complementary in that it probes conditions in the upper layers of the atmosphere. Thus, we present observations primarily from Hinode EIS, IRIS, SDO's HMI and RHESSI, with SDO's AIA (Atmospheric Imaging Assembly) \citep{Lemen2012} providing supporting context. The temporal evolution of the flare as observed by GOES and RHESSI is plotted in the right-hand column of Figure 1, with the vertical lines indicating the time intervals of the sunquake, while on the left is the IRIS Mg II slit-jaw image from the peak of the impulsive phase, illustrating the ribbon structure in the chromosphere. Marked on the slit-jaw image are the location of the IRIS spectrograph slit (vertical line), and the locations of the sunquake (SQ), and of a region within the northern ribbon that we call the non-quake (NQ) region. This region was chosen for comparison because it represented an area of very strong UV and EUV intensity enhancement outside the SQ location, and was scanned by the IRIS spectrograph.

EIS observed the region from 14:05 - 17:57 UT, executing the high cadence flare study PRY\_flare\_1, which covered an FOV of 44$^{\arcsec}\times120^{\arcsec}$ centred on (499$^{\arcsec}$,274$^{\arcsec}$) with the 2$^{\arcsec}$ slit in 4$^{\arcsec}$ steps to produce a raster cadence of 134 seconds. The study included eight line windows, including the He II 256.3 \AA\ (log T$_{e}$=4.7)and Fe XXIII 263.76 \AA\  (log T$_{e}$=7.1) lines, which are the main focus of this work. The EIS data were calibrated using the standard software routine eis\_prep, and the slit tilt and the orbital variation of the line position were corrected for. The lines were fitted with either a single or two-component Gaussian fit, which allowed us to construct intensity, Doppler velocity and width information for each raster.

SDO's HMI and AIA instruments both observed the whole disk of the Sun and the flare throughout its duration. Standard calibration (aia\_prep) and image alignment procedures were applied, and the other data were aligned to these data. Co-alignment with EIS can be challenging given the small FOV, and the temporal smearing that is introduced by the rastering of the slit. Thus, while the offset between EIS and AIA is now routinely recorded, and is better than 5$^{\arcsec}$ in x and y, we confirmed our alignment using feature matching of the integrated intensity in the Fe XII 192.41 \AA\ window and images from the AIA 193 \AA\ bandpass recorded at the mid-point of the EIS raster. This alignment, which we estimate to be good to $\pm$ 2$^{\arcsec}$, then allowed us to align EIS, AIA 1600 \AA\ and the IRIS slit-jaw data.

IRIS observed the event both with the slit-jaw imager and the spectrograph. The spectrograph covered an FOV of 14$^{\arcsec}\times174^{\arcsec}$ centred at (491$^{\arcsec}$,282$^{\arcsec}$) with 8 slit positions at a 9.4 second exposure time, giving a total raster cadence of 72 seconds. The line list included the Mg II h \& k lines at 2803 and 2796 \AA\ respectively, the Si IV 1393.76 and 1402.77 \AA\ lines and the C II 1334.53 and 1335.71 \AA\ lines. The slit-jaw imager provided images in the 1400 \AA\ and 2796 \AA\ bandpasses covering a FOV of 167$^{\arcsec}\times174^{\arcsec}$ from 14:09 -17:54 UT. The IRIS pixel size for both the slit-jaw imager and spectrograph is 0.166$^{\arcsec}$. We primarily used the IRIS level 2 spectrograph data in our analysis, and the level 2 calibration includes corrections for dark current, flat-fielding as well as geometric and wavelength corrections. Although IRIS observed in several wavelength windows, the intensity of the flare was such that many of the spectra were saturated during the peak of the impulsive phase. This inherently limits our analysis and interpretation in a way that the ground-based observations of \cite{judge14} are not subject to, and for this reason we note the importance of combining both ground and space-based data sources when possible. In this work we focus primarily on the Mg II h and k lines, and the Si IV line. Both the Mg II and Si IV do show some saturation during the flare, and we exclude these data from our spectral fitting. However, since these wavelength regions are inaccessible from the ground, these data still provide important information about the flare process, and despite the saturation, substantial changes in the Mg II h and k lines are seen during the flare, including large velocity shifts, line broadening, disappearance of the central reversal and the appearance of the subordinate 3p$^{2}$P -3d$^{2}$D  triplet lines, which offer diagnostic potential for flare heating in the lower atmosphere that has so far been exploited relatively little \citep{FeldDos1977, pereira15}. Both the IRIS and EIS spectrographs were operating in a scanning mode during the event studied, and the time to build a spectral image using rastering is inherently longer than the time taken to produce an image using broad-band filters. However, as can be seen from figures 4 - 8 and 10 - 12, we have still been able to sample spectra at important points during the flare evolution.

RHESSI observed for the entire duration of the flare and the data were used to construct light-curves in a range of energy bins, and to reconstruct images in the 25-50 and 50-100 keV energy ranges. For the image reconstruction we used the PIXON algorithm with a 20 second integration time. Figure 2 shows the location of the reconstructed HXR sources at 25 -50 (purple) and 50 - 100 keV (black) with respect to flare ribbons observed in the IRIS Mg II slit-jaw images.

\begin{figure}
\figurenum{1}
\epsscale{1.0}
\plotone{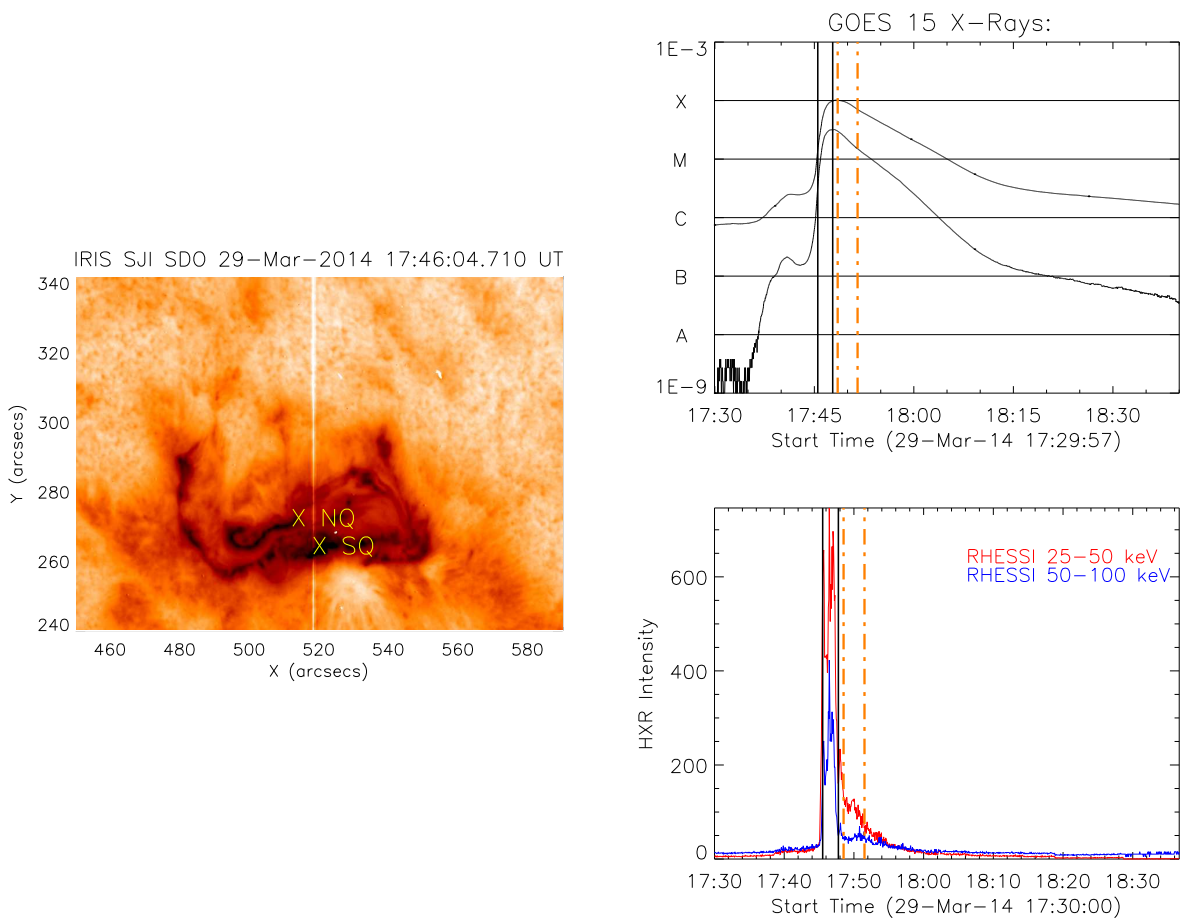}
%\plotone{matthews_fig1.eps}
\caption{Left: IRIS Mg II slit-jaw image showing the location of the sunquake (SQ) and the non-quake (NQ) region and the location of the spectrograph slit (vertical line). Right: top: GOES X-ray intensity for the flare; bottom: RHESSI count-rates in the 12-25 keV (red) and 25-50 keV (blue) energy ranges. The solid back (dashed orange) lines represent the time intervals of strong photospheric Doppler velocities and the derived sunquake onset time, respectively.}
%\ref{fig:context}
\end{figure}

\subsection{Acoustic methods}

As noted in the introduction, the presence of the sunquake has already been reported by \cite{judge14}, and the holography data that we present here are independent confirmation of those results. However, we also include complementary time-distance analysis that confirms the location and timing of the sunquake. For the holography we also used the HMI photospheric data obtained aboard SDO with 45 second cadence \citep{Scherrer2012}. For helioseismic analysis the data are tracked at the Snodgrass differential rotation rate and re-mapped using Postel projection for helioseismic purposes.  In order to detect acoustic emission associated with the flare, we applied both acoustic holography and surface time-distance analysis to the tracked Dopplergram data. The holography method \citep{Donea1999} uses Green's function $G_{+} (|\vect{r}-\vect{r}'|, t-t') $ which prescribes the acoustic wave propagation from a point source, to essentially 'backtrack' the observed surface signal $ \psi(\vect{r}, t)$. This allowed egression images to be reconstructed showing the sub-surface acoustic sources and sinks:
\begin{equation}
H_{+}({\vect{r}, z, t})=\int dt' \int_{a<|\vect{r}-\vect{r}'|<b} d^{2}\vect{r}' G_{+} (|\vect{r}-\vect{r}'|, t-t') \psi(\vect{r}', t'),
\end{equation}
where $a$, $b$ define the holographic pupil. The egression power is then defined as :
\begin{equation}
P(z, \vect{r})=\int dt |H_+({\vect{r}, z, t})|^{2} dt
\end{equation}
where $r'$ and $\theta$ are the polar coordinates describing the vector $|\vect{r}-\vect{r}'|$ at the solar surface.

The time-distance analysis \citep{kz1998, ZH11, Zharkov2013b} uses the surface line-of-sight velocity obtained from the three-dimensional series of observations (two horizontal dimensions plus time) by selecting the source location, re-writing the observed surface velocity signal in polar coordinates relative to the source, then contracting the azimuthal dimension via integration applying a Fourier transform. The result is a time-distance diagram presenting the radial surface distance from the source versus time seen as a ridge following the time-distance relationship for the acoustic waves derived from Dopplergrams, pointing to the circular ripples propagating on the solar surface from the source where a flare occurs.

Photospheric velocity changes are seen in the HMI Doppler data within the southern flare ribbon at $~(519^{\arcsec},262^{\arcsec})$ between 17:45:31 - 17:47:46 UT. During large flares it is likely that HMI channels will be affected by intensity core reversals, and indeed these are seen in the magnetic field data. Also, the photospheric data analysed by \cite{judge14} indicate quite modest velocities in the quake location. Most of the sunquake detections made using acoustic holography reported up until recently have been characterised by a well-defined localised acoustic kernel. Such kernels are normally seen in egression measurements at more than one frequency. While the shape and extent of the kernels may vary from frequency to frequency, they tend to be cospatial. The shape, longevity and prominence of kernels may also depend on the choice of the pupil used in egression. But again, the location of strongest emission remains approximately the same. As Figure 1 in \cite{judge14} shows, and is independently verified by our measurements, the 6mHz egression data in this event are quite noisy, indicating several potential kernels. We also found that the shapes and locations of these kernels varied considerably from frequency to frequency as well as with the choice of pupil. To localise the egression power emission, we used a 15 by 15 Mm integration box to analyse the acoustic emission statistics for all data points at different frequencies, marking the locations where signal passes $5\sigma$ threshold. This, combined with time-distance analysis, allowed us to verify and localise the source. In Figure 3, we show the 6 and 7mHz egression power contours (cyan and magenta, respectively), together with the 25 -50 and 50 -100 keV  HXR contours overlaid on the IRIS 2976 \AA\ slitjaw image at 17:46 UT, indicating that there was strong energy deposition in this location at many wavelengths. 
Figure 2  shows the time-distance plot associated with the acoustic source at  $~(519^{\arcsec},262^{\arcsec})$, and from this we determine an onset time is in the range 17:48:31 - 17:51:31  $\pm$45 seconds. 
This source is also the focus of the \cite{judge14} work, and they report two onset times for the egression sources at  5.5 and 6 mHz of 17:46 and 17:51 UT, respectively. In our analysis we also find some evidence for different evolution at different frequencies, with a peak time between 17:47:46 - 17:50:01 UT at 7 mHz, and between 17:50:46 - 17:51:31 UT at 6 mHz. It should be noted that because of the filtering involved, the uncertainties in onset times derived from holography can be $\pm$5 minutes.

\begin{figure}
\figurenum{2}
\epsscale{0.7}
\plotone{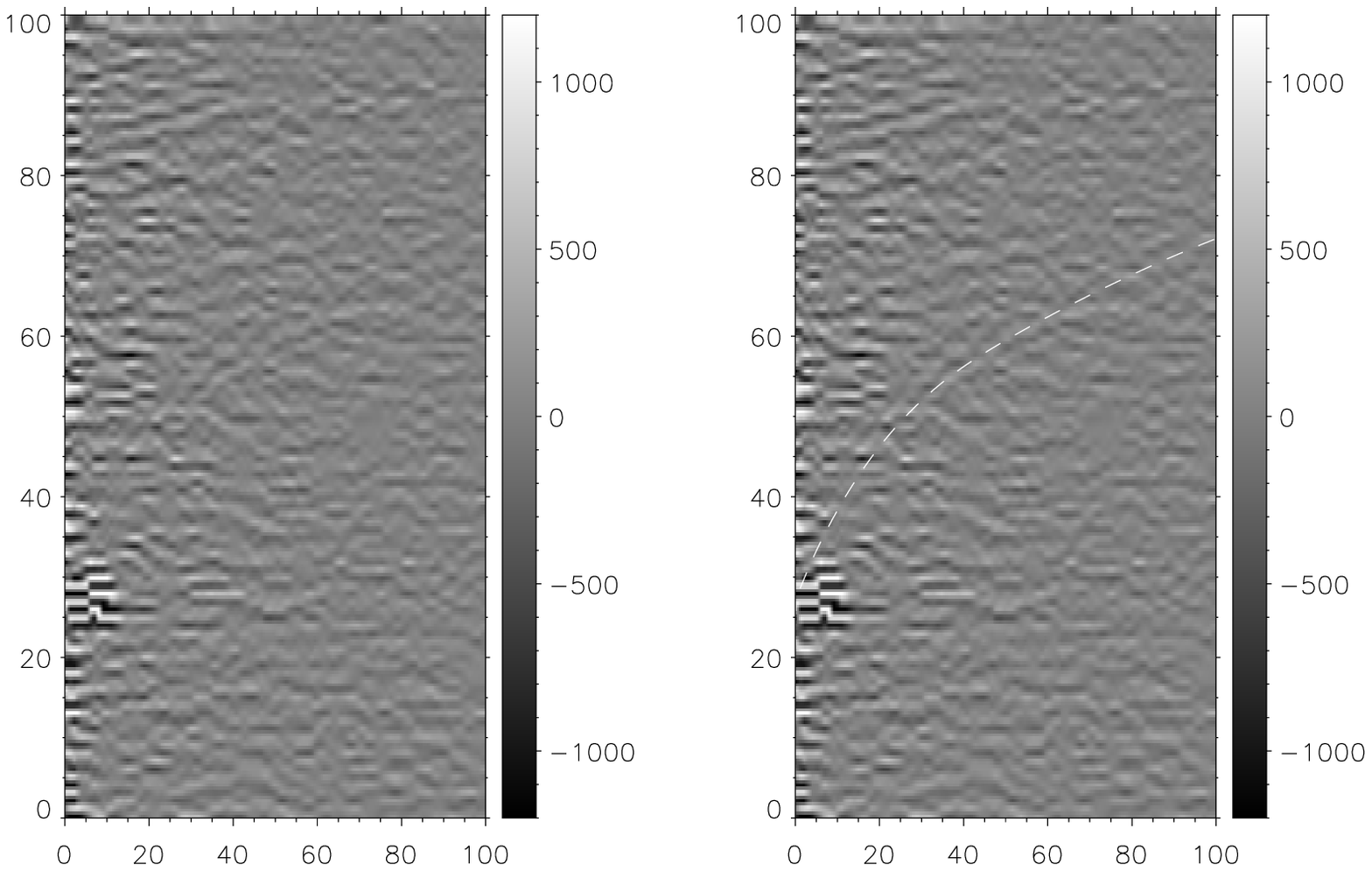}
\caption{Time-distance diagram for the acoustic source at (519$^{\arcsec}$,262$^{\arcsec}$) (left), with the theoretical travel time ridge over plotted (white dashed line) on the right.}
\end{figure}

\section{Results}

The egression source located at (519$^{\arcsec}$,262$^{\arcsec}$) has the greatest statistical significance, and is also the location of the time-distance source.  In order to complement to the work of \cite{judge14} in trying to understand the processes that led to the production of the acoustic disturbances at this particular location of the flaring environment, we use spectroscopy to probe the conditions that prevailed in the overlying corona, transition region and chromosphere at this position, and compare them to conditions at another bright location (NQ) in the northern flare ribbon at (511$^{\arcsec}$,272$^{\arcsec}$). These positions can be seen in Figure 3 to correspond to the sites of strong Mg II emission during the peak of the impulsive phase at approximately 17:46:30 UT, and for the sunquake, to the strongest 6.0 mHz egression contours and HXR emission. The NQ region is at the edge of the strongest HXR emission in the northern ribbon, and this location is at the Eastern edge of the FOV of the IRIS spectrograph, so we were unfortunately unable to study the Mg II spectral response any closer to the centroid of the HXR emission in the Northern ribbon.

The intensity and dynamics of the flare produced complex profiles in all the spectral lines that we examined, as can be seen in figures 4, 5, 6 and 7, where we plot the He II 256.3 \AA\, Fe XXIII 263.76 \AA\, Si IV 1402.8 \AA\ and Mg II  h and k spectra for the NQ location (left column) and SQ location (right panels), respectively. 

\begin{figure}
\figurenum{3}
\epsscale{0.7}
\plotone{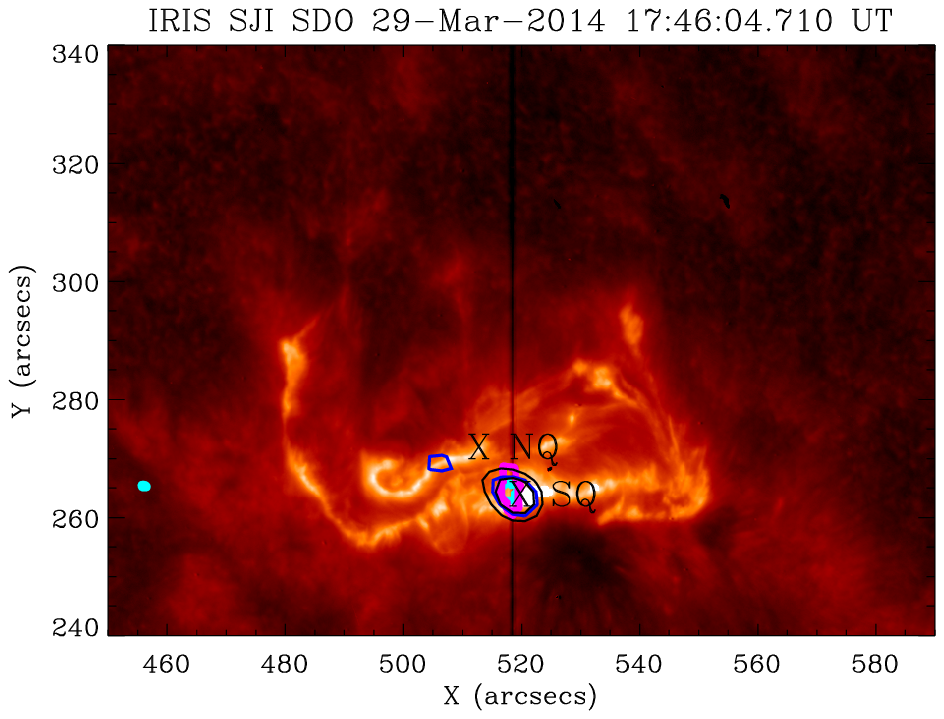}
\caption{ IRIS Mg II slit-jaw image showing the location of the sunquake (SQ) and the non-quake (NQ) region and the location of the spectrograph slit (vertical line), with 6 (cyan) and 7 (magenta) mHz egression contours and RHESSI HXR contours (blue: 25 - 50 keV; black: 50 -100 keV).}
\end{figure}

\subsection{He II spectra}
The He II 256.3 \AA\ spectra were fitted with a combination of single and double Gaussian component fits, and the reference wavelength was derived from an average of the pre-flare spectra to be 256.36 \AA\. In Figures 4 and 5 we plot the He II spectra from the NQ and SQ locations (c.f. Figure 1 for locations within the flare ribbons). Figure 4 shows spectra from the impulsive phases of the flare (14:42 - 17:47 UT), while Figure 5 shows the spectra from the gradual phase (17:48 - 17:55 UT). Since both EIS and IRIS raster in order to build up an image, the timing at each slit location within the image differs by the time taken to move between raster positions; in this case 24 seconds, and so it is important to keep in mind that each location is not measured simultaneously. However, it can be seen from this figure that the He II profile in the SQ location is consistently broader and of greater intensity than in the NQ location, and that it is red-shifted with respect to the reference wavelength, with the greatest red-shift of $\sim$80 kms$^{-1}$ occurring at 17:46:38 UT. It is interesting to note that an extended red 'tail' begins to develop  in the SQ location around 17:51 UT, with a very strong and highly shifted red component apparent at 17:55 UT. The spectra in the NQ location indicate a significant red-shift and red wing component at 17:47 UT, but thereafter the line centroids indicate only small deviations from the reference wavelength, indicating that there are no significant motions of the emitting plasma.

\begin{figure}
\figurenum{4}
\epsscale{0.9}
\plotone{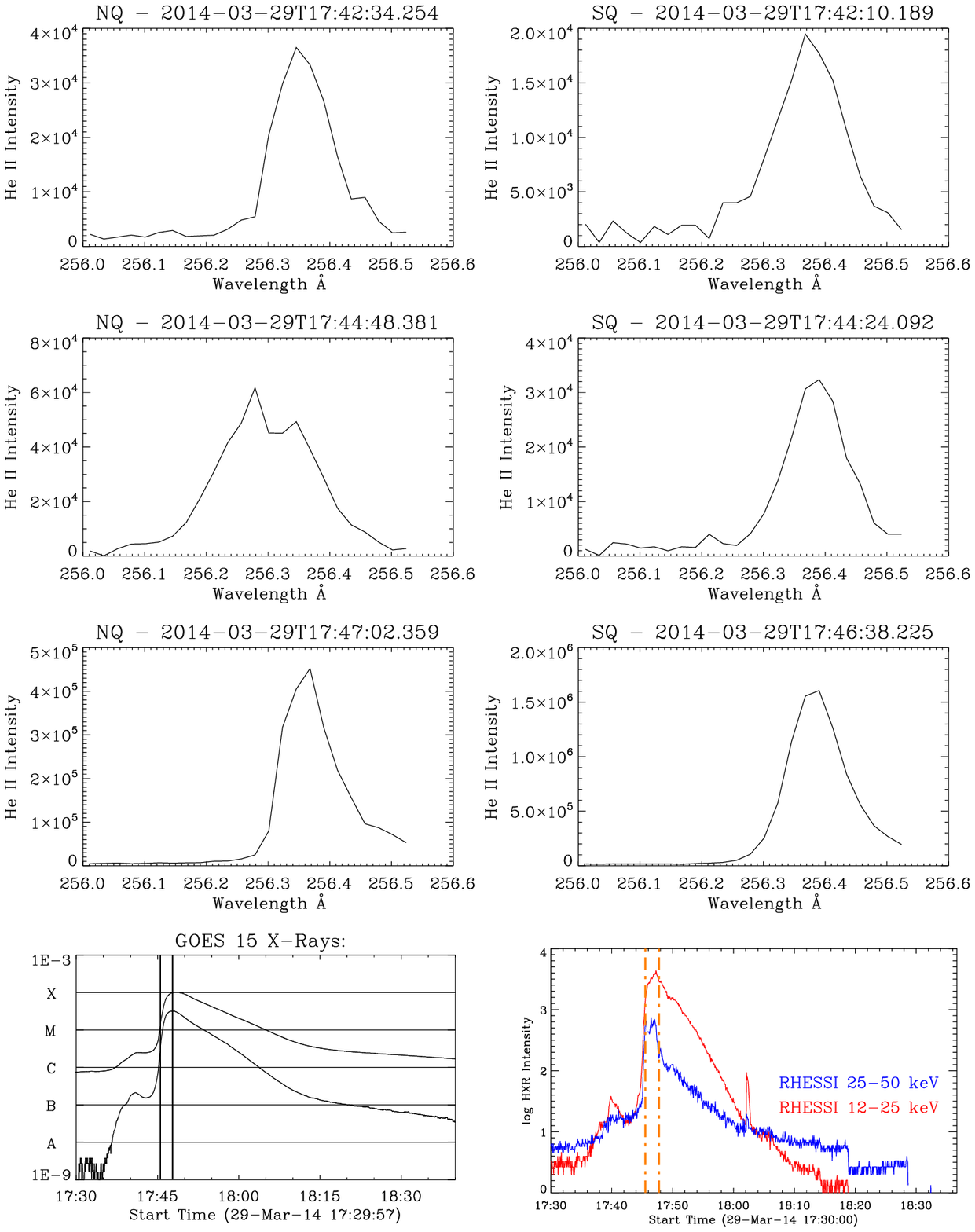}
\caption{EIS He II 256.3 \AA\ spectra at the location of the SQ (right column) and at the NQ location (left column) during the impulsive phase of the flare, as indicated by the vertical lines on the GOES and RHESSI HXR light-curves.}
\end{figure}

\begin{figure}
\figurenum{5}
\epsscale{0.7}
\plotone{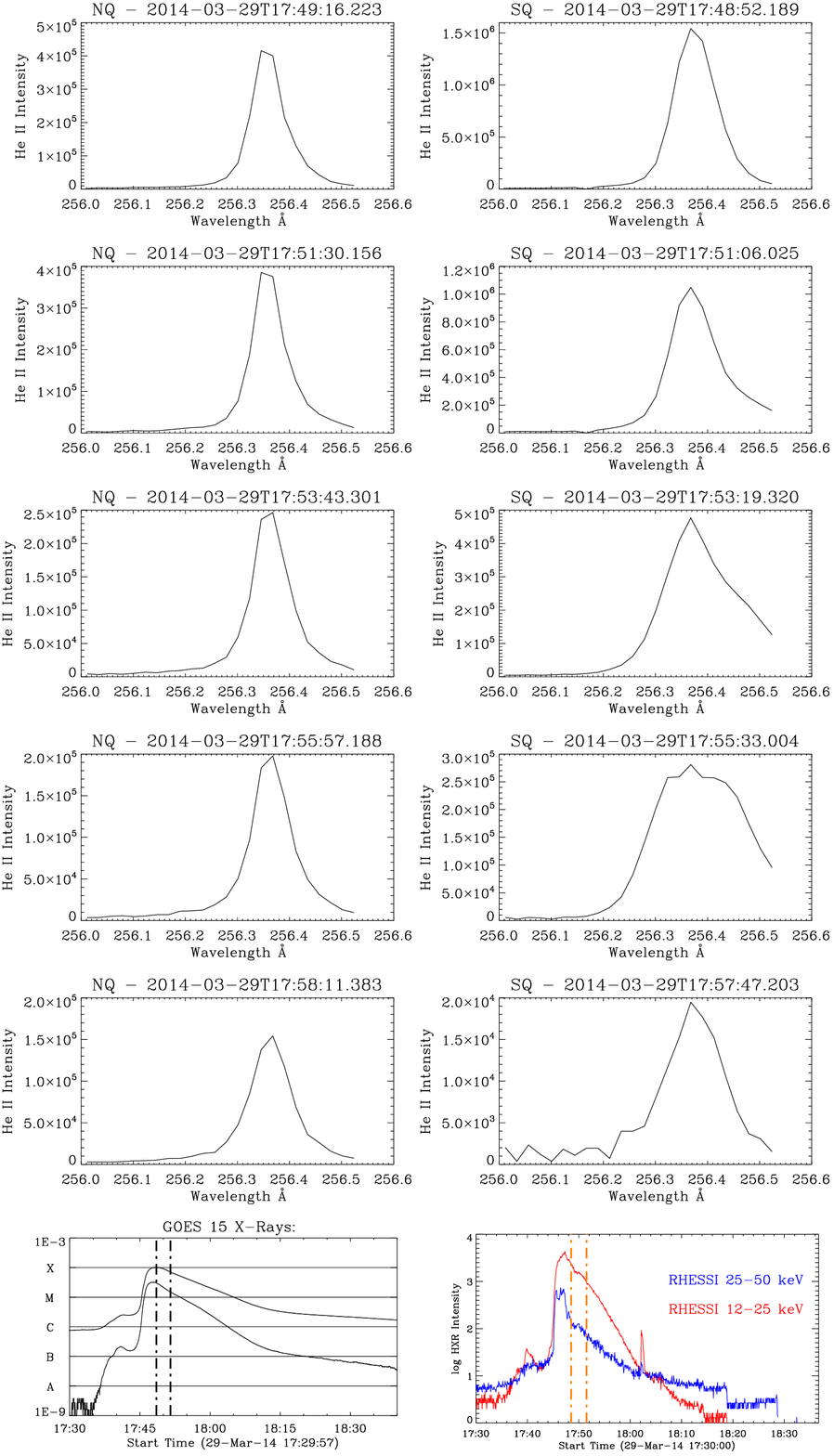}
\caption{EIS He II 256.3 \AA\ spectra at the location of the SQ (right column) and at the NQ location (left column) during the gradual phase of the flare, as indicated by the vertical lines on the GOES and RHESSI HXR light-curves.}.
\end{figure}

\subsection{Fe XXIII spectra}
Prior to 17:46 UT, there is no significant Fe XXIII 263.78 \AA\ emission visible in either the SQ or NQ locations, although it is observed further East in the active region. However, as can be seen from Figure 6, the spectra in both locations show a substantial blue component from 17:46 UT onwards, reaching $\sim$280 kms$^{-1}$ in the SQ location. In the NQ location, low intensity blue wing emission is seen until 17:51 UT, while in the SQ location the profile is symmetric but shifted to longer wavelengths from 17:48 UT onwards. The peak red-shift velocity in the SQ location is 26 kms$^{-1}$ at 17:51 UT.

\begin{figure}
\figurenum{6}
\epsscale{0.7}
\plotone{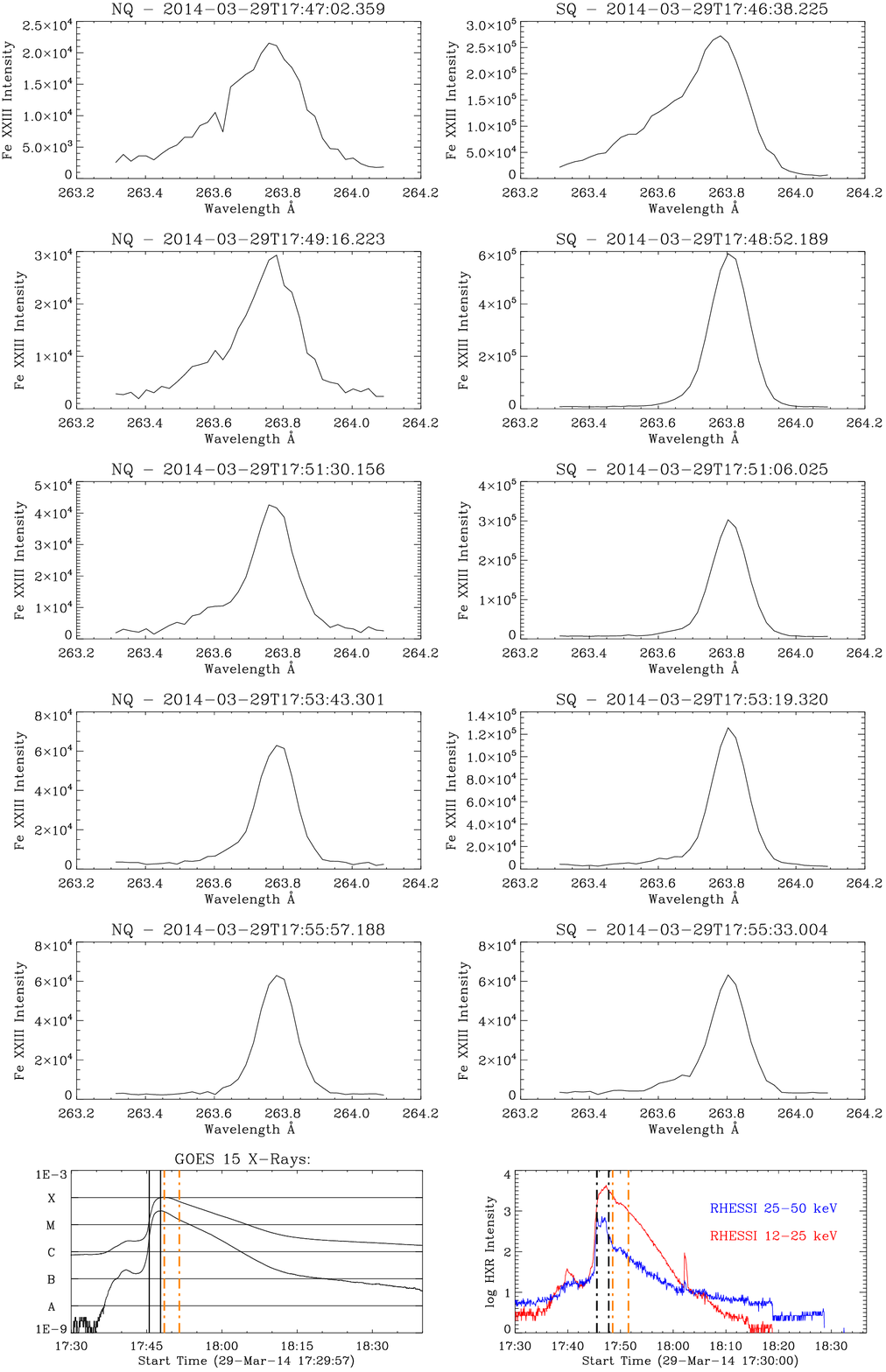}
\caption{EIS Fe XXIII 263.7 \AA\ spectra at the location of the SQ (right column) and at the NQ location (left column) during the impulsive (black lines) and gradual phases (orange lines) of the flare, as indicated by the vertical ones on the GOES and RHESSI HXR light-curves.}
\end{figure}

\subsection{Si IV spectra}
From pre-flare spectra we derive a reference wavelength of 1402.82 \AA\ for the Si IV line. As can be seen from Figures 7 and 8, where we plot the NQ and SQ spectra together with reference light-curves from GOES and RHESSI, it is clear that both the NQ and SQ locations suffer from saturation at a number of times during the flare impulsive phase (the third panel down on the right in figure 7 is blank for this reason).  However, the NQ location shows strongly blue-shifted components at 17:43 and 17:44 UT, probably associated with the onset of the filament eruption \citep{kleint15}, followed by saturation during the flare peak (17:45-17:46 UT), and the subsequent development of a red 'tail' from 17:48 UT onwards. In the SQ location, there is an initial red-shift at 17:43 UT, which develops into an extended red tail at 17:44 UT. The spectra are saturated at 17:46 and 17:47 UT, and we then see the development of a strong red component from 17:48 UT, with a peak downward velocity of 86 kms$^{-1}$ at 17:52 UT. In Figure 9, we plot the intensity and Doppler widths derived from the Si IV line fits, omitting the saturated spectra. While the intensity and Doppler width at the flare peak are not accurately reflected in these plots, what is clear is that there are multiple peaks in the SQ location, with the first occurring before the impulsive phase peak, at $\sim$17:43 UT, and a later strong burst after the impulsive peak at $\sim$17:51 - 17:53 UT. A substantial increase in Doppler width is also seen coincident with the later intensity increase.

\begin{figure}
\figurenum{7}
\epsscale{1.0}
\plotone{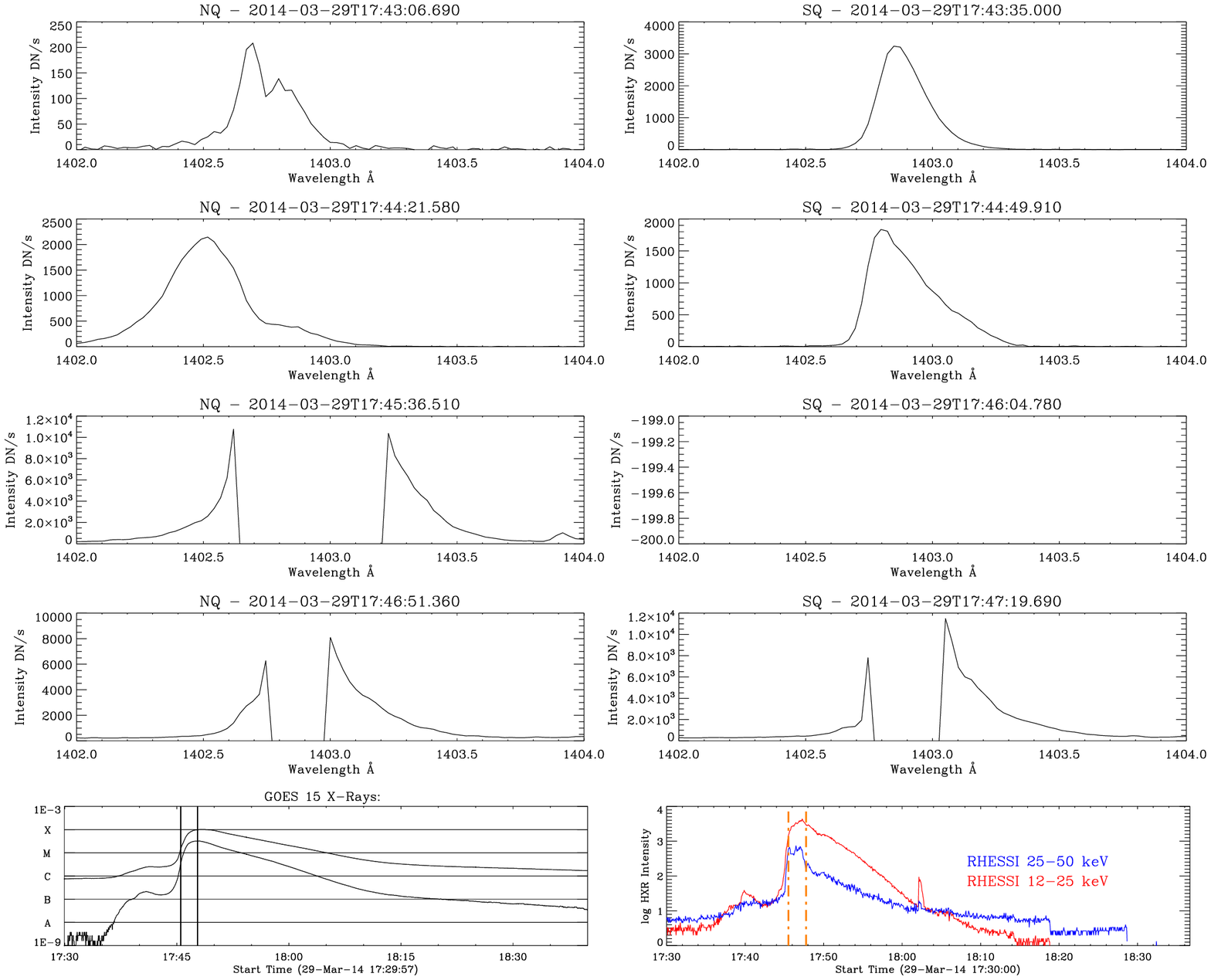}
\caption{IRIS Si IV  \AA\ 1403 \AA\ spectra at the location of the SQ (right column) and at the NQ location (left column) during the impulsive phase of the flare, as indicated by the vertical lines on the GOES and RHESSI HXR light-curves.}
\end{figure}

\begin{figure}
\figurenum{8}
\epsscale{0.8}
\plotone{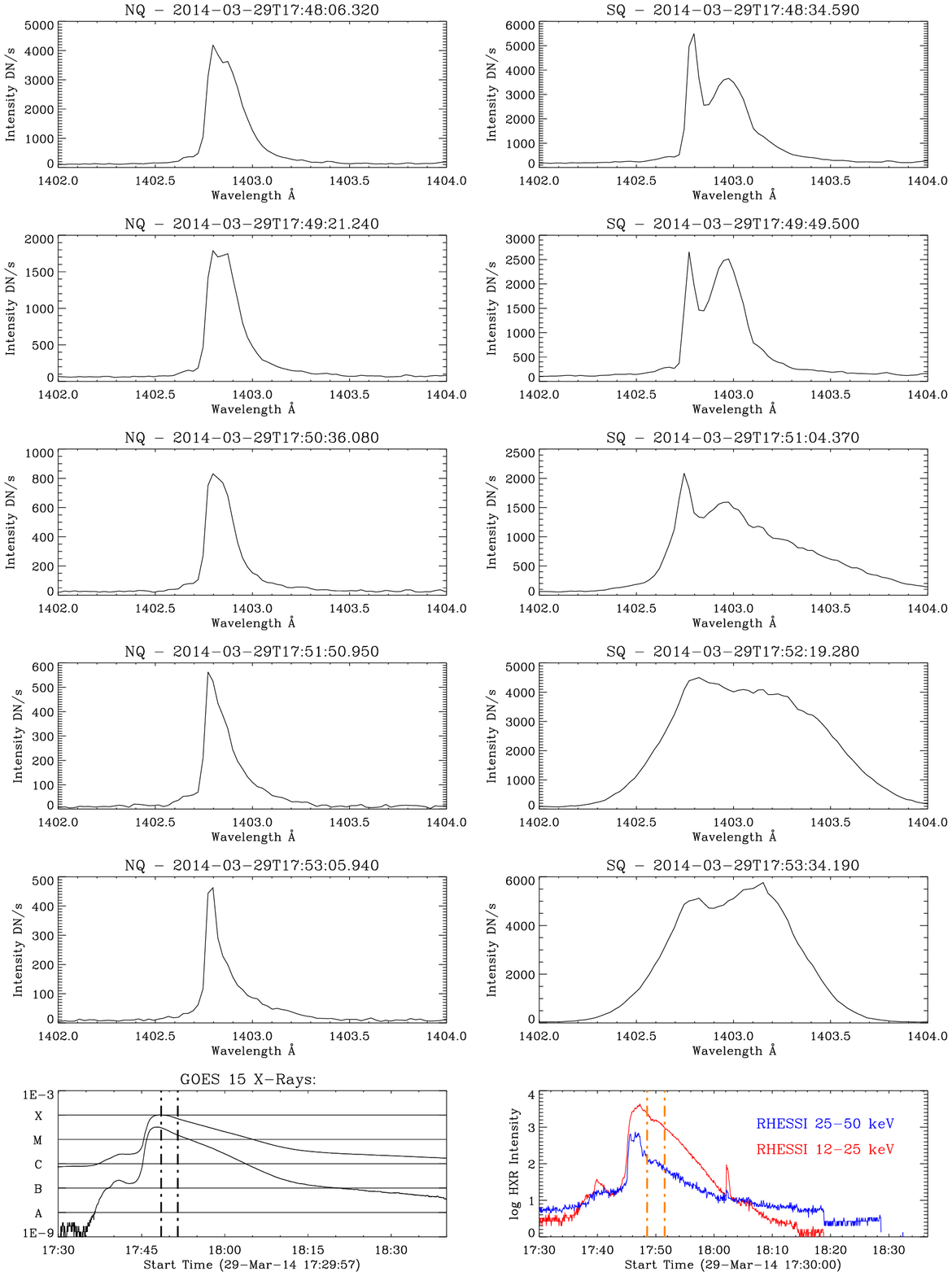}
\caption{IRIS Si IV  \AA\ 1403 \AA\ spectra at the location of the SQ (right column) and at the NQ location (left column) during the gradual phase of the flare, as indicated by the vertical lines on the GOES and RHESSI HXR light-curves.}
\end{figure}

\begin{figure}
\figurenum{9}
\epsscale{0.9}
\plotone{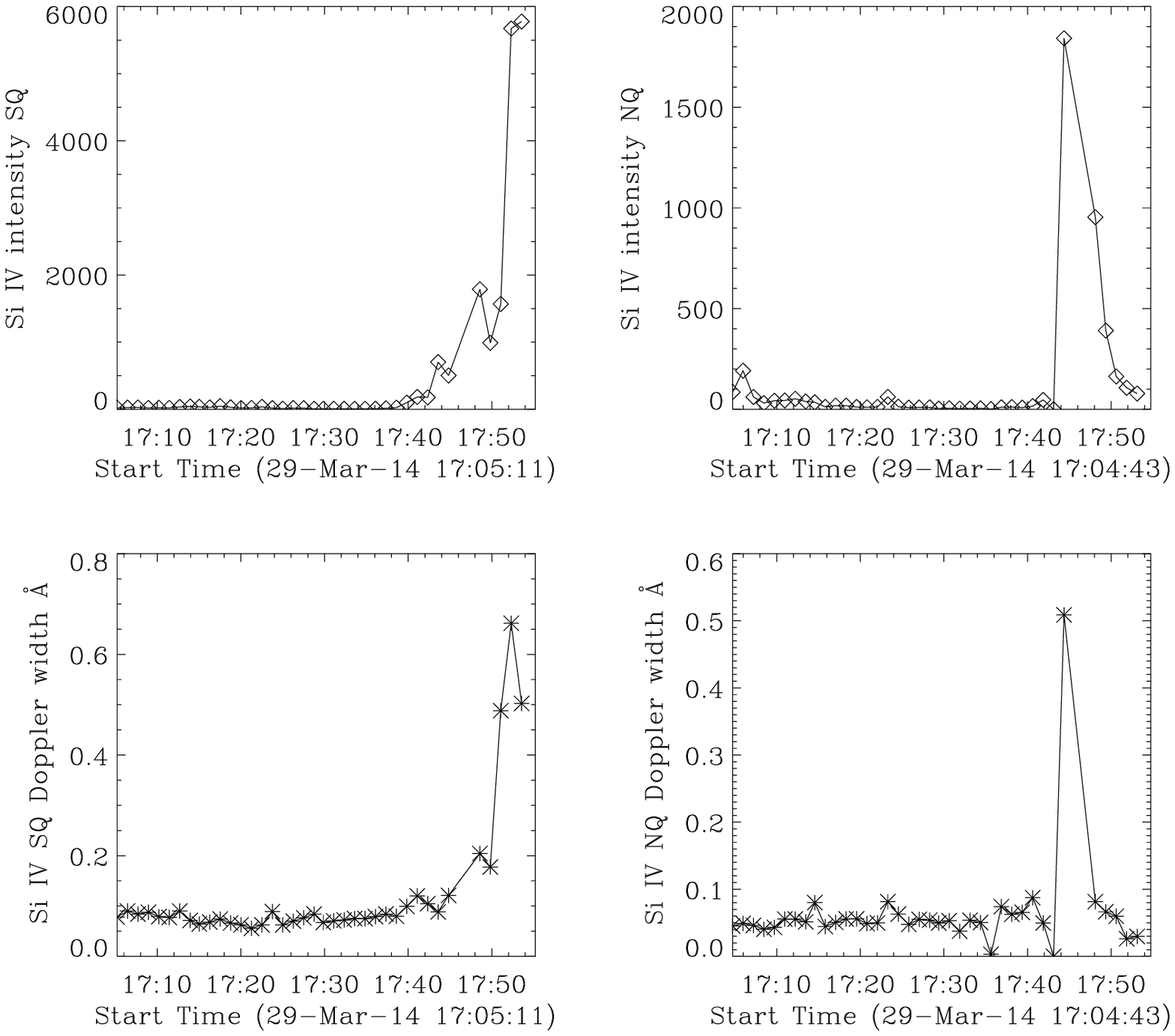}
\caption{Top: Si IV intensities derived from spectral fitting for the SQ (left) and NQ (right) locations. Bottom: Doppler widths for the SQ (left) and NQ (right) locations}
\end{figure}

\subsection{Mg II h and k spectra}

As for the Si IV line, the Mg II  spectra are heavily saturated in the location identified with the sunquake, as can be seen in the third panel on the right in Figure 10. This made it impossible to fit the spectra and derive the intensity or velocity at this position. However, we were able to fit the spectra in the SQ location in preceding and successive rasters (Figure 11), as well as the spectra in the NQ location. It should be noted that the Mg II and Si IV lines can be optically thick, and hence fits to the spectra do not simply reflect plasma properties, but do provide a means to comment on relative changes in the line centroid and profile. The spectra in both locations showed complex variations in the line profiles of both the h and k lines, as well as in the profiles of the companion triplet lines that arise from the 3p$^{2}$P and 3d$^{2}$D transitions at 2791.60 \AA\ and 2798.75 and 2798.82 \AA\ . In particular we note the disappearance in the central reversal, the increase in line width that can be observed in both the NQ and SQ locations at the time of the flare peak, the increase in intensity and width of the triplet lines, and the subsequent narrowing and decrease in intensity of both the resonance and triplet lines in the NQ location following the flare peak. In the SQ location similar behaviour is observed until 17:51 UT, when there is evidence of the development of a red component in the h and k lines, and a decrease in the intensity of the triplet lines. The subsequent rasters at 17:52 and 17:53 UT then show very broad h and k profiles, and much reduced peak intensity in the triplet lines, although these are also broad. The 'jagged' nature of the h and k lines is puzzling, and may be due to an instrumental effect. However, the observed intensities at this time are well below saturation levels. 

Using the fits to the Mg II h and k lines, we derived the k/h intensity ratio at both the SQ and NQ locations from 17:30 - 17:53 UT, as shown in Figure 12. From this figure we can see that the k/h ratio in the SQ location (left column, second panel) shows little variation prior to flare onset, maintaining a value of $\sim$1.1, but then begins to rise at flare onset (17:44 UT), peaking around 17:50 UT and then falling to pre-flare values. Note that because of the saturation this plot does not include an estimate of the ratio at the flare peak at 17:46 UT, nor does the top panel include an estimate of the intensity at this time. In the NQ location (right column), the pre-flare ratio is similar, with a maximum of 1.37 observed at flare onset (17:44 UT), followed by a subsequent drop to a minimum at 17:45:36 UT, and then a gradual return to pre-flare levels. In the optically thin limit one would expect this ratio to be 2, and the ratio at all times prior to and during the flare remains less than 2, indicating that the lines remain optically thick. The presence of the saturation makes it impossible to comment on the behaviour of the ratio at the peak of the impulsive phase in the SQ region, but it is intriguing that the ratio decreases significantly compared to pre- and post-flare values during the period 17:44 -17:49 UT in the NQ region, but does not appear to do so in the SQ region.

We also measured changes in the Doppler widths at the SQ and NQ locations in both the h and k lines, and these are also plotted in Figure 12; again, SQ in the left-hand column and NQ in the right-hand column. In this case we have used the extent of the saturation at flare maximum to make an estimate of the line width at this time, but this should only be viewed as an upper limit. For both locations it can be seen that there is little variation in line width prior to flare onset, and the widths are very similar in both locations. A substantial increase in line width is seen at flare onset and peak in both cases, but even exercising due caution because of the saturation, it is clear from the spectra that the h and k line widths at the SQ location are at least 50\% larger those seen in the NQ location, reaching $\sim$1.6 \AA\. As can also seen in the spectra in figure 7, the h and k line widths then increase again in the SQ location at $\sim$17:50 UT. The saturation in the h line is less than for the k line, so we use this line to derive an upper limit of 1.833 \AA\ for the line width at 17:46:04 UT. Subtracting the instrumental and thermal widths, this corresponds to a non-thermal velocity of 190.1 kms$^{-1}$. If we assume that this broadening represents a flux of Alfv\'en waves, then following \cite{judge14} we can estimate the associated wave energy density:

\begin{equation}
F_{wave}=\rho v_{A}\langle v_{nt}^{2} \rangle \unit{erg \; cm^{-2} \; s^{-1}}
\end{equation}

where $\rho$ is the mass density, $v_{A}$ is the Alfv\'en speed, and $v_{nt}$ is the velocity amplitude of the Alfv\'en wave. \cite{pereira15} find that the Mg II triplet lines are sensitive to column densities in excess of 5 $\times$10$^{-4}$ g cm$^{-2}$, so since we are unable to directly measure the density, $\rho$, we use the value tabulated by \cite{VAL} for a column mass of 7.82$\times$10$^{-4}$ g cm$^{-2}$, which corresponds to a mass density of 5.49 $\times$10$^{-4}$ g cm$^{-3}$. Using B = 800 G, as measured by \cite{judge14}, we calculate the energy flux density to be 6.04$\times$10$^{11}$ erg s$^{-1}$ cm$^{-2}$. \cite{judge14} calculate a lower limit for the peak of the acoustic power to be  5$\times$10$^{9}$ erg s$^{-1}$ cm$^{-2}$. We stress that since the spectra are saturated, and this may artificially enhance the line width, the energy flux derived at this time is an upper limit, and we also note the large ($\sim$140 G )uncertainty on the magnetic field strength measurement. However, the spectra at 17:52 UT are not subject to saturation, and a similar calculation yields a non-thermal velocity of 152.2 kms$^{-1}$ and associated wave energy flux of 3.87 $\times$10$^{11}$ erg s$^{-1}$ cm$^{-2}$ at this time. The error on the magnetic field strength measurement is not large enough to reduce either of these values by two orders of magnitude.

The saturation in the resonance lines in the SQ location at the flare peak (17:46 UT) makes it impossible for us to derive a Doppler velocity at that time, although it is apparent from Figures 10 and 11 that at 17:46, 17:52 and 17:53 UT these lines are either red-shifted or have a substantial red wing component. In fact, fits to the k line at 17:52 UT indicate a red component with a velocity of $>$ 100 kms$^{-1}$. The 2798.75 and 2798.82 \AA\  triplet pair are also saturated, at 17:46 UT, but the line at 2791.60 \AA\  is free from saturation, and so we are able to estimate velocity for this line. A substantial red component is seen that peaks at 54 kms$^{-1}$, with a width of 0.899 \AA\ that causes it to blend with the Fe I line at 2792.394 \AA\ , which also displays red-shift of 54 kms$^{-1}$. 

In Figure 13 we plot the relationship between the emission at 17:46 UT (left) and 17:52 UT (right) from different temperature plasmas with respect to the egression power at 6 and 7 mHz derived from acoustic holography. The background image in both cases (reverse red colour table) is the emission in the blue wing of Fe XXIII at 280 kms$^{-1}$. On this background image we then plot the 6 and 7 mHz egression contours in cyan/magenta for both times; the 25-50 and 50-100 keV HXR emission contours at 80 and 90\% of the maximum (purple and black, respectively) at 17:46 UT, and the 25-50 keV (purple) contours  at 17:52 UT; emission from He II at 60 and 80 kms$^{-1}$ (magenta) at both times, and emission from the red wing of the Mg II 2791.6 \AA\ triplet line at 70 kms$^{-1}$ at both times.

Recent work by \cite{pereira15} highlights the diagnostic potential of the Mg II triplet lines, finding that the presence of these lines in emission requires steep temperature increases in the lower chromosphere ($>$ 1500 K) and typical electron densities  of 10$^{18}$ m$^{-3}$ or greater. They also find that the ratio of the core to wing intensity is correlated with the temperature increase required to drive the lines into emission, and that the shape of the emission line profile provides further information about the temperature profile. Their simulations indicate that multiple peaked profiles are generally the result of a temperature profile that varies rapidly with height, while lines with a single peak are consistent with a single, strong temperature increase at higher altitudes. When there is a strong temperature increase low in the atmosphere, and at high density, then their simulations predict that the emission will be seen in the far wings of the line, but that the line core will show a central reversal, and the resonance lines will be very broad. \cite{pereira15} demonstrate the correlation between core to wing intensity and temperature increase for the long wavelength pair of the triplet lines. For the 29 March 2014 flare, this line pair is saturated at the flare peak. However, we can derive the 2791.6 \AA\ core to wing ratio, and we find this to be 2.21 for the SQ location, which comparing to Figure 5 of \cite{pereira15}, indicates a temperature increase T(S$_{max}$) - T$_{min}$ $\simeq$ 2000 K, where T(S$_{max}$) is the temperature at the height where the source function maximum occurs, and T$_{min}$ is the temperature minimum between column masses of 10 - 0.1 g cm$^{-2}$. The profile of the triplet line at 2791.6 \AA\ shows an asymmetric central reversal, with a strong enhancement in the red wing, while the h and k lines are clearly strongly broadened, suggesting that the temperature increase in this case occurring deep in the atmosphere, and at high densities. An alternative explanation for the increased intensity of the Mg II triplet lines may be found in the work of \cite{judge05} and \cite{laming92} who discuss the effects of intense shorts bursts of heating that create 'ionising' plasma conditions, whereby there is a strong burst of emission in the metastable transition as the result of more effective population of these levels.

\begin{figure}
\figurenum{10}
\epsscale{1.0}
\plotone{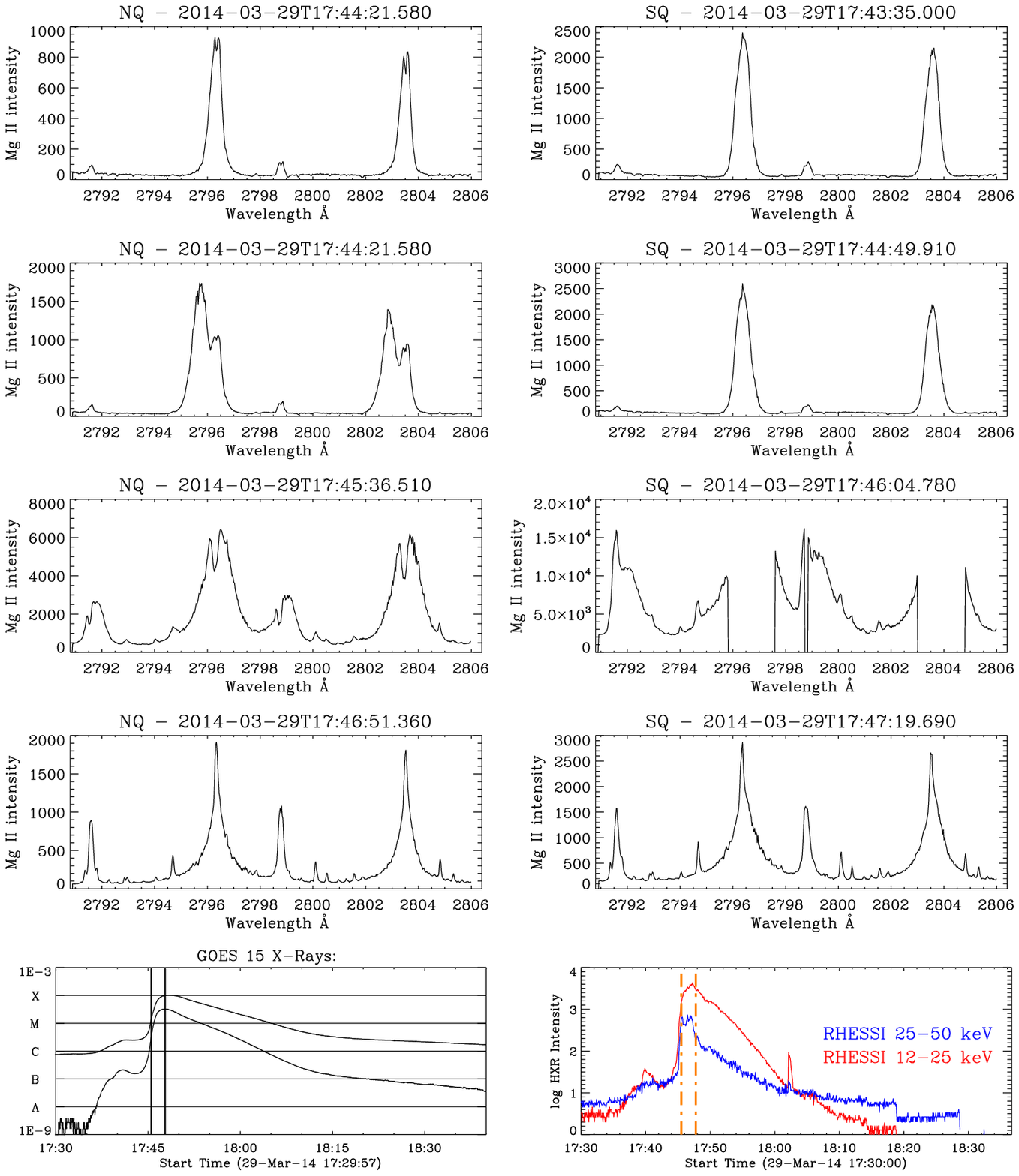}
\caption{Mg II k and h spectra at the location of the SQ (right column) and at the NQ location in the North flare ribbon (left column) during the impulsive phase of the flare, as indicated by the vertical lines on the GOES and RHESSI HXR light-curves.}
\end{figure}

\begin{figure}
\figurenum{11}
\epsscale{0.75}
\plotone{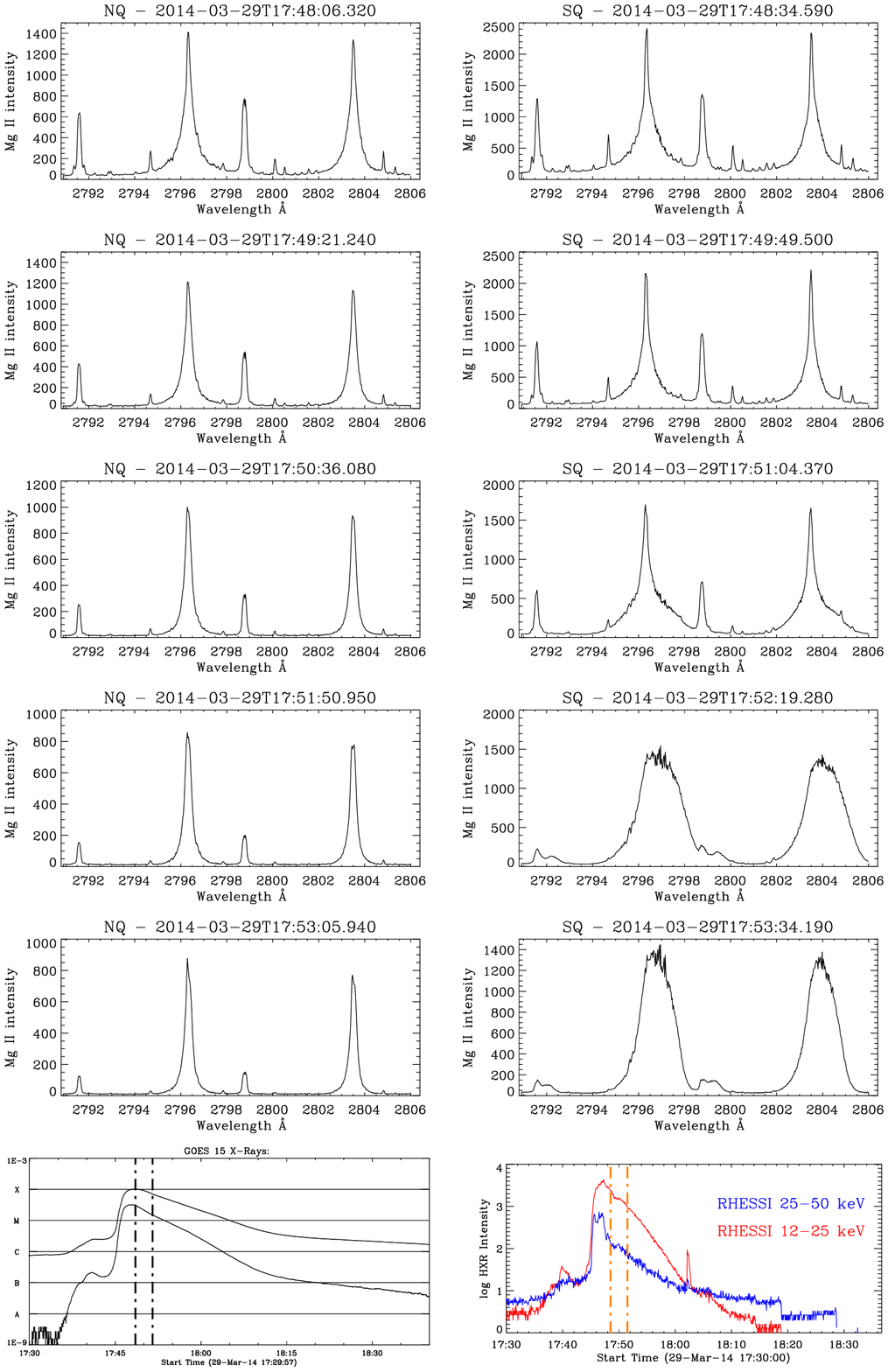}
\caption{Mg II k and h spectra at the location of the SQ (right column) and at the NQ location in the North flare ribbon (left column) during the gradual phase of the flare, as indicated by the vertical lines on the GOES and RHESSI HXR light-curves.}
\end{figure}

\begin{figure}
\figurenum{12}
\epsscale{0.7}
\plotone{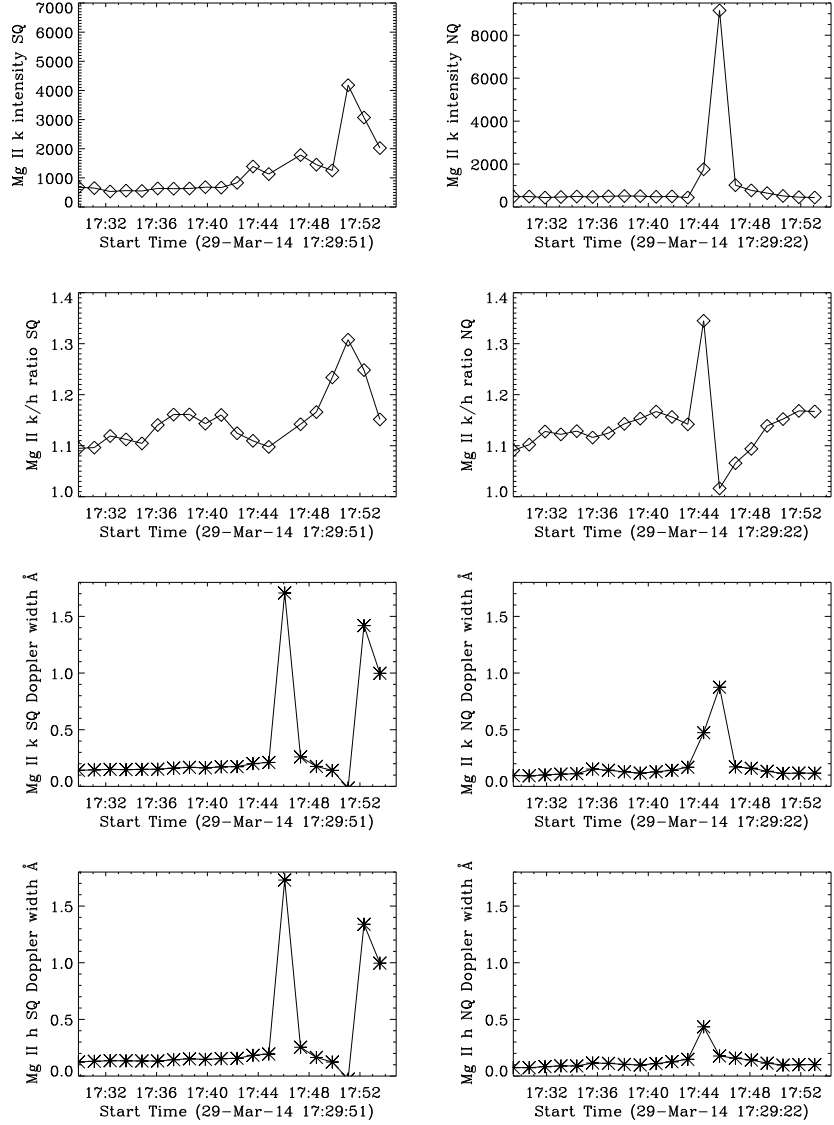}
%\plotone{matthews_fig12.eps}
\caption{Left: from top to bottom: Mg II k intensity; k/h ratio; Mg II k and h Doppler widths for the SQ location. Right: from top to bottom: Mg II k intensity;  k/h ratio; Mg II k and h Doppler widths for the NQ location.}
\end{figure}

\begin{figure}
\figurenum{13}
\epsscale{1.2}
\plottwo{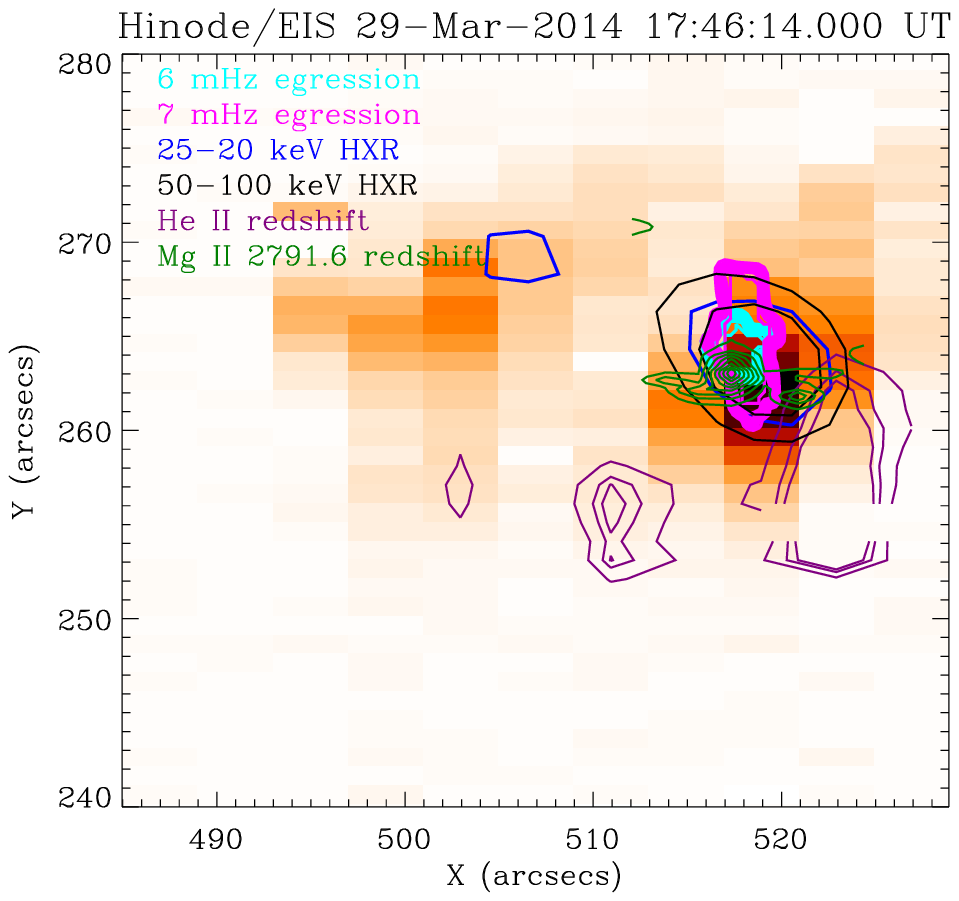}{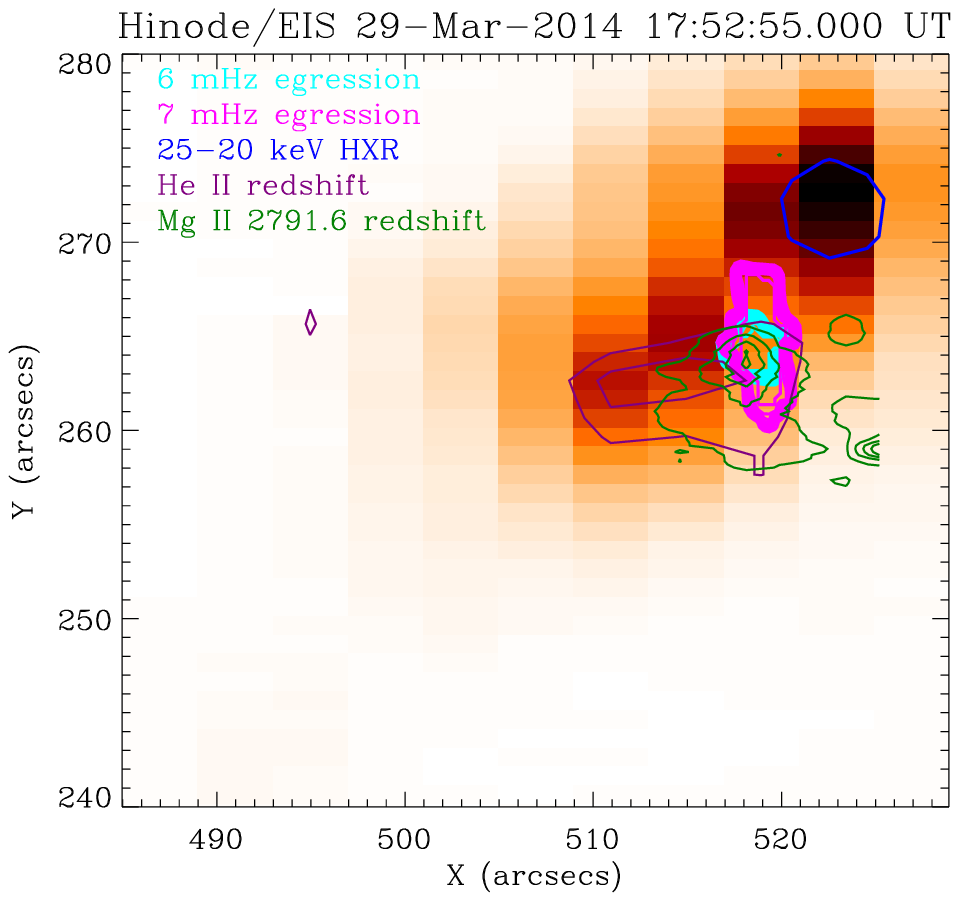}
\caption{Left: Background image: Fe XXIII blue wing emission at 280 kms$^{-1}$; 6 and 7 mHz egression contours in cyan/thick magenta; 25-50 and 50-100 keV HXR emission contours at 80 and 90\% of the maximum (purple and black, respectively); emission from He II at 60 and 80 kms$^{-1}$ (magenta), and emission from the red wing of the Mg II 2791.6 \AA\ triplet line at 70 kms$^{-1}$ (green). Right: as left, but at 17:52 UT.}
\end{figure}

\section{Discussion and Conclusions}
We report new spectral observations of the chromosphere, transition region and corona from Hinode EIS and IRIS above the location of the sunquake detected during the X1 flare of 29 March 2014, which complement the observations and analysis already done by \cite{judge14} on this event. We compare the spectra from the upper atmosphere above the sunquake with spectra from another location in the flaring ribbons that exhibits strong emission at multiple wavelengths, but which has no associated acoustic disturbance. Using holography and time-distance methods, we confirm the existence of an acoustic impact in the photosphere and below, as reported by \cite{judge14}. The time-distance analysis indicates an onset time between 17:48:31 - 17:51:31 UT $\pm$45 seconds, while the holography suggests some differences in evolution with frequency, with peak times between 17:47:46 - 17:50:01 UT at 7 mHz, and between 17:50:46 - 17:51:31 UT at 6 mHz, which are also consistent with the findings of \cite{judge14}. During the time of the 7 mHz peak signal we find that in the location of the quake there are spatially correlated upflows ($\sim$280 kms$^{-1}$) in the hot Fe XXIII line, intense HXR emission in the 50 - 100 keV energy range, downflows of 80 kms$^{-1}$ in the He II line, strongly broadened emission in the Mg II h and k lines ($\sim$1.78 \AA\ ) and intense and broad emission in the Mg II triplet lines, including a red wing component with velocities in the range $\sim$ 54 - 70 kms$^{-1}$. In contrast, in the NQ location we find that, while the Mg II h and k and triplet lines are intense and broadened with respect to pre-flare values, the intensity increase is an order of magnitude less, and the increase in line width is around a factor of two smaller. At the time of the 6 mHz peak signal ($\sim$ 17:51 UT), while the GOES and integrated HXR data (see Fig. 1) indicate a likely second burst of energy release, we find no increase in Mg II or Si IV intensity, line width or Doppler velocity in the NQ location, but at the SQ location we find an increase in Mg II  and Si IV intensity, strongly broadened  Si IV and Mg II h and k profiles ($\sim$1.5 \AA\ ), and a red-shifted component of $>$ 100 kms$^{-1}$ in the Mg II lines. However, in this time interval while we observe the Mg II triplet lines to be broadened, the intensity is two orders of magnitude less intense than at 17:46 UT, and we observe no measurable HXR emission at 25 - 50 or 50 - 100 keV in either the SQ or NQ location. 

We computed the intensity ratio of the Mg II k to h lines in both locations, omitting the saturated spectra in the quake location at 17:46 UT. We find this ratio to be significantly less than 2 for both locations, before and during the flare, but with a significant impulsive increase at flare onset, followed by a sharp decrease. In the SQ location the ratio is comparable to pre-flare values at the time of the 6 mHz peak, while it is significantly lower in the NQ location. Comparing our observations to the simulations and diagnostics presented in \cite{pereira15} we find the that profile of the Mg II triplet line at 2791.6 \AA\ and the ratio of the wing/core intensity in this line are consistent with a single, strong increase in temperature of $\sim$2000 K deep in the atmosphere in the location of the quake. However, we note that the increase in intensity of these lines is also consistent with the ionising plasma condition described by \cite{judge05}.
 
\cite{Wolff72} first suggested that solar flares may drive acoustic disturbances in the solar interior. The subsequent detection of the first quake by \cite{kz1998} not only confirmed these predictions, but suggested that the driver of the acoustic disturbance was a strong compression shock generated as the result of sudden and rapid heating of the chromosphere within the framework of the 'thick-target' model e.g. \citep{Brown71, kostiuk75}. In this scenario, the rapid heating and compression drive a radiative instability that forces the heated chromospheric plasma upwards (chromospheric evaporation), while generating an opposite downward propagating shock, which then drives the acoustic disturbance. In this picture, the heating and subsequent evaporation are driven by the non-thermal electrons that are generated by the reconnection process in the corona.

The majority of sunquakes discovered prior to the launch of SDO showed a good spatial and temporal correspondence not only with hard X-ray signatures, but also with enhancements in the white-light, leading researchers to also consider the role of pressure transients that are related to back-warming of the photosphere by the enhanced chromospheric radiation \citep{Donea1999, DL2005}, as well as the direct precipitation of electron (and proton) beams \citep{Zharkova2002}. However, more recently observed events \citep{alvarardo12}, and statistical studies \citep{pedram12} have indicated that the site of strongest white-light emission is not always coincident with the quake location. More recently, the role of the Lorentz force transients that occur whilst the coronal field is being reconfigured \citep{Hudson_etal12}, have been considered as another possible driver. Within this context one must also consider the fact that non-thermal particle beams carry a strong electric field \citep{Zharkova06}, that in turn can induce an electromagnetic field in the ambient plasma \citep{Zharkova11} that modifies the magnetic field of the loop where particles precipitate. Magnetic field changes of this type are attractive as a potential driver since they do not require the same delay that the propagation of shocks requires, and there is increasing evidence that many sunquakes are associated with erupting flux ropes (e.g. \cite{Zharkov_etal11}; Green et al., (2015), in preparation). However, so far it has proved difficult to identify a definitive relationship between changes measured in the photospheric magnetic field during the flare, and the site of the acoustic emission. In part, this may in fact be due to effects of the flare energy deposition, which are known to produce artefacts in the magnetic field data \citep{martinez14}, but in relation to the event studied here, \cite{judge14} estimate that the energy flux associated with the observed changes in the photospheric magnetic field are insufficient to account for the observed acoustic power of the sunquake. It is worthy of note that the sunquake in this event is somewhat unusual in that, while it is located in field that is of similar strength to the surrounding penumbra, there is no penumbra visible in this location in the continuum data. The majority of other quakes have been found in penumbral regions. 

Our investigations of the 29 March 2014 sunquake have focussed on spectroscopic signatures of plasma conditions in the overlying chromospheric, transition region and coronal layers, and their comparison with another location in the flare that displays no acoustic emission. These observations are complementary to those reported by \cite{judge14} which provide important insights into the photospheric environment around the sunquake, where we would expect to see significant energy deposition. Our holographic analysis indicates that the sunquake associated with this event is somewhat unusual in displaying a quite dispersed signal, but the spatial and temporal coincidence of the strongest egression kernels seen at 6 and 7 mHz and the time-distance source is compelling evidence for its reality. We find evidence for some variation in properties of the quake at different frequencies, with an earlier peak seen at 7 mHz than at 6 mHz. This first peak is coincident with the rise and peak of the flare impulsive phase, as observed in HXR between 17:45:31 - 17:47:46 UT (Fig. 1), while the second occurs during the decay of the HXR burst between 17:48:31 - 17:51:31 UT. As can be seen from Fig. 13, at 17:46 UT we observe a very close spatial coincidence between strong HXR emission at 25 - 50 and 50 -100 keV, an intense, high velocity blue-shifted component in the hot Fe XXIII line ($\sim$280 kms$^{-1}$), strong red-shifted emission ($\sim$70 kms$^{-1}$) in the Mg II  2791.6 \AA\ triplet line and 7 mHz acoustic emission. The He II 256.3 \AA\ line also shows down flows, but slightly offset from the other emission. The profile of the Mg II 2791.6 \AA\ triplet line, and the ratio of the core/wing intensity also appear consistent with simulations by \cite{pereira15} that suggest a single strong temperature increase of  2000 K deep in the atmosphere. Taken together, these observations seem to strongly suggest the driver of the acoustic impact in the interval 17:47:46 - 17:50:01 UT is consistent with a thick-target scenario, where rapid heating by non-thermal electrons drives chromospheric evaporation and an opposite downward propagating shock that excites the sunquake. On the other hand, the later 6 mHz peak in the interval $\sim$17:51 UT seems to have a different driver, related to a second energy release at the SQ location at this time. In this case we observe no measurable HXR emission in the 25 - 50 and 50 - 100 keV range at the SQ location, but we observe very broad and red-shifted emission in all the Mg II lines ($>$ 100 kms$^{-1}$ in the k line), He II ($\sim$80 kms$^{-1}$), Si IV ($\sim$86 kms$^{-1}$) and a small red-shift of $\sim$26 kms$^{-1}$ in hot (log T=7.1) Fe XXIII emission. The lack of HXR emission in the SQ location (this true at lower energies as well) is inconsistent in this case with the thick-target scenario, but the combination of very broad, red-shifted spectral profiles throughout the atmosphere may be consistent with the presence of a significant flux of Alfv\'en waves. Using the measured line widths in the Mg II h and k lines we estimated the wave energy flux that would be associated with the presence of Alfv\'en waves, and we compare this with the estimate of the peak acoustic power derived by \cite{judge14} of 5 $\times$ 10$^{9}$ erg s$^{-1}$ cm$^{-2}$, both at 17:46 and 17:52 UT, finding values of 6.04$\times$10$^{11}$ erg s$^{-1}$ cm$^{-2}$, and 3.87 $\times$10$^{11}$ erg s$^{-1}$ cm$^{-2}$ respectively. We stress that the impact of saturation at the earlier time, and our inability to actually measure the density, introduce significant uncertainties into these estimates, but we think it unlikely that these would be two orders of magnitude, suggesting that Alfv\'en waves may indeed play a role in driving the sunquake in this event. 

We note that it is puzzling that we observe an apparently very strong chromospheric response to the sunquake, while there appears to be a relatively weak response observed in the photosphere, particularly when the observations presented by \cite{judge14} were so fortuitously located in space and time. We have no satisfactory answer to why this might be at this stage, but speculate that perhaps the wave energy flux that we observe (assuming for now that that is the source of the line broadening) dissipates between the regions of formation of the Mg II and Si I emission. What that might imply for the mechanism(s) of sunquake generation requires further investigation. Finally, we want to emphasise the complexity of the event studied here, which on the one hand seems to strongly support a role for chromospheric shocks in driving the acoustic disturbance, but also indicates the likely presence of a strong flux of Alfv\'en waves. Coupled with the recent results that indicate an inconsistent correlation of sunquakes with the brightest white-light flare emission, and an association with the erupting flux rope, this highlights the difficulties in identifying a single driver, if indeed there is only one.

%% The displaymath environment will produce the same sort of equation as
%% the equation environment, except that the equation will not be numbered
%% by LaTeX.


%% Included in this acknowledgments section are examples of the
%% AASTeX hypertext markup commands. Use \url without the optional [HREF]
%% argument when you want to print the url directly in the text. Otherwise,
%% use either \url or \anchor, with the HREF as the first argument and the
%% text to be printed in the second.

\acknowledgments
The research leading to these results has received funding from the European Commission's Seventh Framework Programme under the grant agreement No. 284461 (eHEROES project).
We would like to thank the referee, Phil Judge, for his thorough and insightful comments that have greatly improved the clarity of the paper.

\bibliographystyle{aa}
\bibliography{lang-5}

\clearpage

%% the file, uncomment the \plotone, \plottwo, and
%% If you need a layout that cannot be achieved with \plotone or
%% \plottwo, you can invoke the graphicx package directly with the
%% \includegraphics command or use \plotfiddle. For more information,
%% please see the tutorial on "Using Electronic Art with AASTeX" in the
%% documentation section at the AASTeX Web site,
%% http://www.journals.uchicago.edu/AAS/AASTeX.
%%
%% The examples below also include sample markup for submission of
%% supplemental electronic materials. As always, be sure to check
%% the instructions to authors for the journal you are submitting to
%% for specific submissions guidelines as they vary from

%% This example uses \plotone to include an EPS file scaled to
%% 80% of its natural size with \epsscale. Its caption
%% has been written to indicate that additional figure parts will be
%% available in the electronic journal.

\end{document}